\documentclass{llncs}
\usepackage{subfigure}
\usepackage[hyphens]{url}
\usepackage{graphicx}

\newcommand{\figwidth}{0.47}
\newcommand{\figwidthtriple}{0.47}

\begin{document}

\title{What slows you down?\\ Your network or your device?}
%\title{Should I upgrade my smartphone?}

\author{Moritz Steiner \and Ruomei Gao}

\institute{Akamai Technologies, San Francisco\\
\email{\{moritz,rgao\}@akamai.com}}

\maketitle
\begin{abstract}
This study takes a close look at mobile web performance. The two main parameters determining web page load time are the network speed and the computing power of the end-user device. Based on data from real users, this paper quantifies the relative importance of network and device. The findings suggest that increased processing power of latest generation smart phones and optimized browsers have a significant impact on web performance; up to 56\% reduction in median page load time from one generation to the following. The cellular networks, on the other hand, have become so mature that the median page load time on one fiber-to-the-home network (using wifi for the last meter) is only 18-28\% faster than cellular and the median page load time on one DSL network is 19\% slower compared to a well-deployed cellular network. 

\end{abstract}

\section{Introduction}
Poor web site performance makes customers abandon transactions, i.e. a customer does not buy a pair of shoes because of frustration with poor web experience. A study from 2014 shows that only 51\% of mobile users are willing to wait for more than two seconds~\cite{userstudy}. Therefore it is important to understand how fast a website is --- most importantly, if it is fast enough. This might depend on many factors, such as the network to which a user is connected, the device used, and of course the website itself. We are especially interested in the performance of cellular networks and mobile devices. Both traditionally have a reputation of being slow~\cite{wang2011web,nikravesh2014mobile}.

Traditionally, web performance was passively measured by the speed of object delivery. Every website consists of a large number of objects, and if the server logs indicate that the objects were delivered in a timely manner, the assumption was made that the website loaded quickly on a user's device. However, server logs do not record the transfer time of an object, and more importantly, server logs do not record how long it took the user device to parse the content, to render the website, and to paint the screen. Moreover, typically, no single server delivers all the objects that make up a site; in fact they are pulled together from many domains.

An active approach is to use geographically distributed probes that download a list of objects at regular intervals. This method has the same drawbacks as the log line analysis --- it does not reveal the true user experience. Later, real browsers were used at the probes to browse the full websites. This was a step in the right direction. However, the deployment of the probes is limited, and so is the number and types of networks they cover. Typically they only use a few different browsers and types of hardware. The bottom line is that using probes one cannot expect to capture the full variety of the various end user experiences. 

In recent years, another attempt was made to measure the user experience. Instead of recording the delivery speed of single objects, browser events get recorded. The Navigation Timing specification~\cite{navtiming} by the W3C defines an interface to query the browser for the timings of various events. In this paper we are going to focus on the loadEvent. This event fired by the browser indicates when a website has finished loading and rendering. 
%Note that some actions can be deferred until after this event, e.g. loading of images below the fold or java-script for user tracking.

A website operator, or a content distribution network (CDN) as a surrogate, can use javascript and the interface provided by Navigation Timing to record the timestamps of interesting events and beacon them back to some data store.  With this technique, the user experience from real users gets measured, which gives it the name Real User Monitoring. Unlike using probes, all the situations of user experience, e.g. bad coverage, and all the scenarios, such as a rarely used browser on an old device, get covered. In contrast to web server log lines only recording when objects have been delivered, this technique takes into account all the processing on the client side needed to render a website. 

In this paper we use a data set of website load timing records to analyze the impact of the device performance and the network performance on the page load time (PLT). Our analysis shows that the performance of the network on the PLT is less significant than the performance of the device. As an example, using a fiber-to-the-home (FTTH) connection (with wifi for the last meters) as opposed to a cellular network speeds up the PLT of a chosen website by 18-28\% in median, but using a later generation phone, e.g. the Nexus 5 instead of the Nexus 4 will speed up the PLT by 24\% or the Galaxy S4 instead of the Galaxy S3 by 30\%. From one iOS version to the next, the median PLT improved by up to 56\%.
%therefore we assume that iOS version correlates with hardware version of the iPhone.

The remainder of this paper is structured as follows. We present some related work~(Sec.~\ref{sec:relwork}) and describe the data collection methodology in more detail~(Sec.~\ref{sec:method}) and give an overview of the data set used in this paper~(Sec.~\ref{sec:data set}). We analyze the impact of device performance~(Sec.~\ref{sec:device}) and of network performance~(Sec.~\ref{sec:network}) on PLT. We discuss the shortcomings of the used methodology and hint possible ways to overcome them(Sec.~\ref{sec:discussion}). Finally, we conclude the paper~(Sec.~\ref{sec:conclusion}).

\section{Data Collection Methodology}
\label{sec:method}

In this paper we measure the web performance as it is experienced by real world users. Analyzing log lines such as HTTP web server log lines or even TCP logs is not sufficient to obtain an accurate picture. It is possible to extract when objects have been requested, how long it took for the web server to start serving these objects, and how long it took to deliver them. The time the client needs to render the site using the received objects cannot be inferred from HTTP or TCP logs. 

The user experience is not so much defined by the transfer times of single objects but by the period of time a user needs to wait before the screen is painted and the site is interactive. Both the network quality and the server are impacting the transfer time, as well as the performance of the device used for rendering, are crucial.

\begin{figure}[t]
\centering
\includegraphics[width=0.99\textwidth]{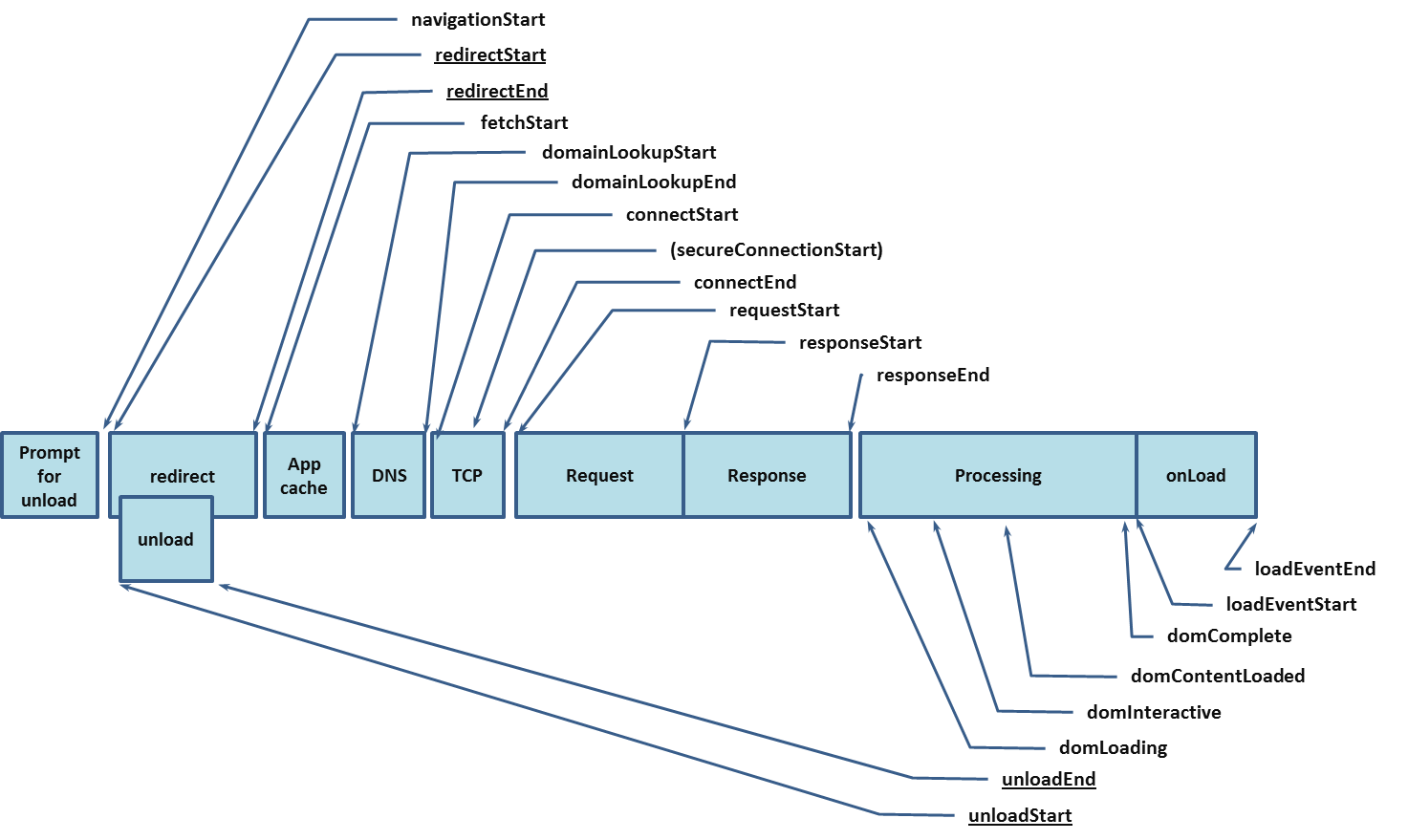}
\caption{Overview of the timing attributes defined in the Navigation Timing specification~\cite{navtiming}.}
\label{fig:timing-overview}
\end{figure}

The Navigation Timing specification~\cite{navtiming} tackles this issue. It defines an interface to access timing information related to web browsing, and is implemented by most major browsers, e.g. Chrome, Firefox, and Internet Explorer. Attributes for the start of the navigation, the domain look up, the connection to the web server, the reception of the response, the building of the document object model, and finally the load event are introduced~(Fig.~\ref{fig:timing-overview}). Real-world performance of sites as seen by real users, on real hardware, and across a wide variety of different networks can be observed using this interface~\cite{meenan2013fast}. In this study we focus on the load event that is fired by the browser when the is loaded and interactive. 
%Note that tricks can be played to delay some actions, e.g. the load of below the fold images or the execution of some java-script, till after the load event has fired. 

As an example on how to use the interface provided, the following code snippet calculates the PLT as the time period between the navigation start event and the load event:

\begin{verbatim}performance.timing.loadEventStart -
performance.timing.navigationStart;\end{verbatim}

For browsers that have not implemented Navigation Timing (e.g. Safari) javascript can be used to obtain the timings of the browsing events manually. Note that the timings obtained through manual javascript as opposed to leverage the Navigation Timing API can't be directly compared.

The full workflow works as follows. We arranged for a web server surrogate (i.e. a CDN proxy server) to insert a javascript snippet into the HTML file requested that recorded the timestamps (with millisecond resolution) of the relevant browsing events and communicates them to a data store back-end. The timing data was enriched with other data, such as the user-agent. From the user-agent, the browser, and in some cases the device type, can be inferred; e.g. a Chrome browser running on Linux, or a Safari mobile browser running on iOS on a iPhone, or a Chrome mobile browser running on a Samsung Galaxy S5. The IP address of the user revealed in combination with an appropriate database not only the general location of the user, but also the network name and the network type (e.g. cellular, cable, DSL, fiber) being used subject to some technical caveats. 

We can analyze the collected data by time, geography, browser, and platform to answer questions such as: What subset of end users are experiencing problems? Is network A faster than network B? Is the latest generation of the smartphone from vendor C really faster than it's predecessor? The performance impact of different versions of the sites themselves or different delivery methods (e.g. HTTP vs. HTTPS, HTTPS vs. SPDY vs. H2, or IPv4 vs. IPv6) can also be analyzed.

Navigation Timing only applies to web content consumed by web browsers. Several major use cases are not covered, e.g. it can't be used for performance analysis of the delivery of file downloads, streaming audio or video, and API traffic triggered by smartphone applications. Other systems and methodologies exist to provide analogous tracking for those use cases, which are not considered in this paper

Note that the PLT is not a universal metric as is throughput. One cannot make simple statements such as "network A is faster than network B" since the PLT is also affected by the device performance. On the other hand one cannot say "device C is slower than device B" if the fraction of those devices on the existing networks is different. And finally comparing the PLT of sites in general is not possible either. In order to make a statement about one of the factors (network, device, and site), it needs to be isolated from the two remaining ones.

\section{Data set}
\label{sec:data set}

The data set used in the paper has been provided by a major CDN operator. Content providers leveraging that CDN can request to have their sites instrumented with javascript for performance timing purposes (see Section~\ref{sec:method}). Typically only a small sample, between less than 1\% up to 5\%, of the sites are randomly selected for instrumentation. Sampling is done to minimize the performance impact due to the javascript insertion on the user.

The data set spans the month of March 2015 and contains 2.5 billions entries of which  9\% come from cellular networks and  28\% are from mobile devices (iPhones and Android devices). Requests are listed coming from all countries and almost 50 thousand AS'es (of which 4\% are listed as cellular) for 92 million distinct URLs (details in Table~\ref{table:data set}).

\begin{table}
\centering
\caption{Overview of the data set used.}
\label{table:data set}
\begin{tabular}{|l|r|} 
%\hline
%\textbf{Country}&\textbf{Fraction}\\ 
\hline
Total number of samples	        & 2,504,914,329\\
Samples from cellular networks	& 214,472,473\\
Samples from mobile devices	& 701,602,020\\
\hline
Samples of URL used for device &  \\  performance analysis (Sec.~\ref{sec:device})    & 1,885,609\\
Samples of URL used for network & \\  performance analysis (Sec.~\ref{sec:network})   & \\
\textemdash{} in the US                       & 500,135\\
\textemdash{} in France                       & 30,287 \\
\textemdash{} country comparison              & 297,039\\
\hline
Distinct URLs                   & 92,065,564\\
Number of ASes                  & 49,913\\
Number of cellular ASes         & 1,655\\
\hline
\end{tabular}
\end{table}

The distribution of PLT across all countries, network types, devices types, and sites is not very revealing. Too many different factors play a role and no useful conclusion may be drawn. In this study, we put the focus especially on cellular networks and mobile devices (i.e. smartphones).

\begin{figure*}[ht]
     \centering
       \subfigure[Boxplot of the PLTs per country.]
        {
           \includegraphics[width=\figwidth\textwidth]{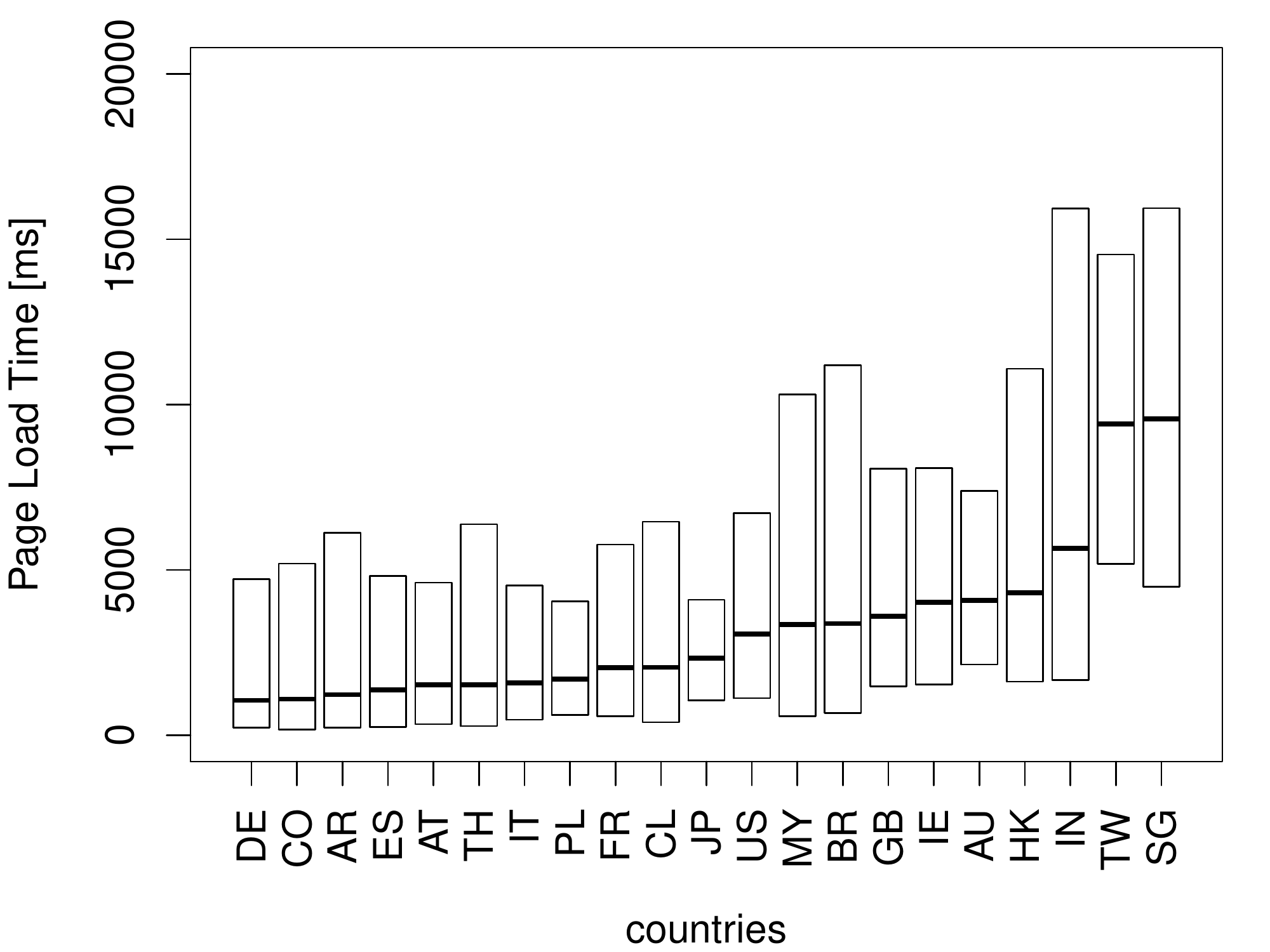}
          \label{fig:topcountry}
          }
      \subfigure[Boxplot of the PLTs per site.]
        {
           \includegraphics[width=\figwidth\textwidth]{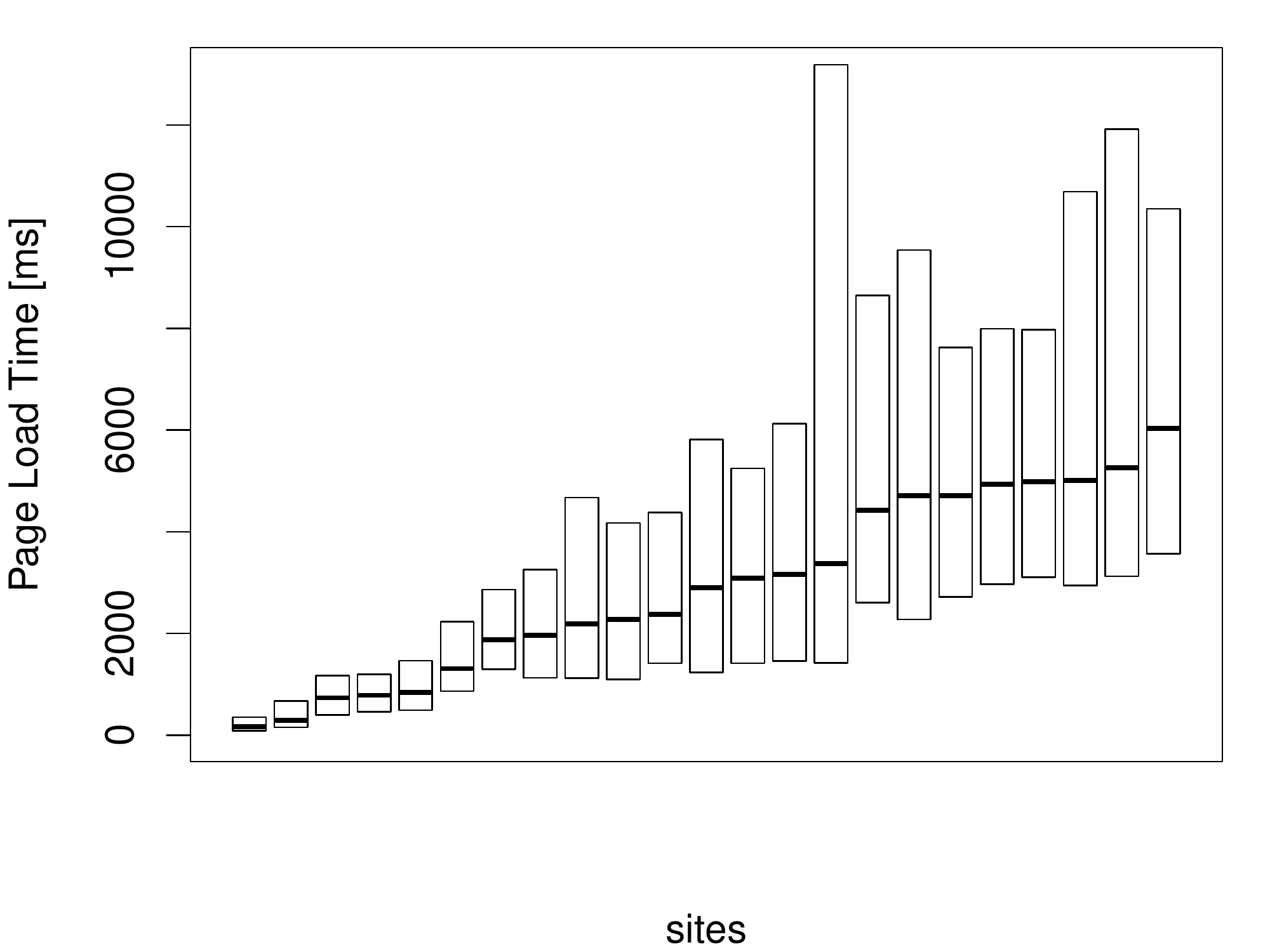}
          \label{fig:topuri}
         }
     \caption{Distributions of the PLTs experienced by mobile devices on cellular networks. The bold line indicates the median and the edges the 25th/75th percentiles.}
     %\label{fig:mice}
\end{figure*}

To start we compare the distribution of PLTs for iPhones and Android devices using cellular networks among the countries with the most entries in the data set~(Figure~\ref{fig:topcountry}). Huge discrepancies can be observed, but why is it that Japans shows a median almost twice as high as Argentina or Columbia? The reason is not that the networks or the devices in Japan are slow, but that the websites visited are heavier and more complex.

Figure~\ref{fig:topuri} shows the distribution of PLTs for the top URLs. The median PLTs vary from a few hundred milliseconds to several seconds. This observation suggests that countries, networks, or devices can't be compared across all pages, but that a single page needs to be picked to allow for a meaningful comparison.

\section{Impact of device performance on the page load time}
\label{sec:device}

In this section we are going to analyze the performance of devices on the PLT. To normalize possible dissimilar browsing behavior, a single URL is picked for the remainder of this section. The URL is an e-commerce landing page optimized for mobile comprising 134 objects with 881 KB in total. Some of the products displayed did change during the observation period but the performance relevant charachteristics remained stable. The data set contains more than 10 million samples for this particular URL.

The performance of the networks included in the analysis could introduce another bias in the comparison. To limit this bias we will confirm in Section~\ref{sec:network} that the performance of the major cellular networks in the US is sufficiently similar to allow for a fair comparison. 
%The latest generation phones might be used mostly in very fast (and expensive) networks versus the older phones might be more prevalent in slower networks (e.g. in countries with less LTE coverage).
The analysis is consequently limited to those networks, reducing the number of samples used to 1.8 million. 

Reductions in PLTs of 18-30\% can be observed from one generation of Android based phones to the following generation. For iOS the user-agent does not allow to identify the hardware version but only the iOS version used. Here the biggest improvement between consecutive versions is 56\%. 
The actual PLTs can't be compared directly between iOS devices running Safari and Android devices running Chrome Mobile, since the later implement Navigation Timing but Safari does not (see Section~\ref{sec:method}).

Devices have a very significant impact on web performance. Unfortunately it is not possible to compare the web performance of smartphones to laptops or tablets using this data set, because smartphones often receive simplified mobile version of a site, while laptops and some tablets receive the full-blown version that typically consists of many more objects, larger images, and more javascript.

\subsection{iPhones}
The user-agent string used by iPhones does not contain information about which hardware (e.g. iPhone 4, 5, or 6) is used, but only the iOS version (e.g. 7, 7.1, 8, 8.1). It seems to be a legitimate assumption to make that there is a correlation between the iOS version and the version of the hardware, since a newly purchased phone will always run the latest version of iOS but not all older phones are updated by their users. The performance gains are partly due to hardware improvements, partly due to software improvements.  With the data collected it is not possible to tell which fraction can be attributed to hardware or software, respectively.  

\begin{figure*}[ht]
     \centering
        \subfigure[iOS versions]
        {
           \includegraphics[width=\figwidth\textwidth]{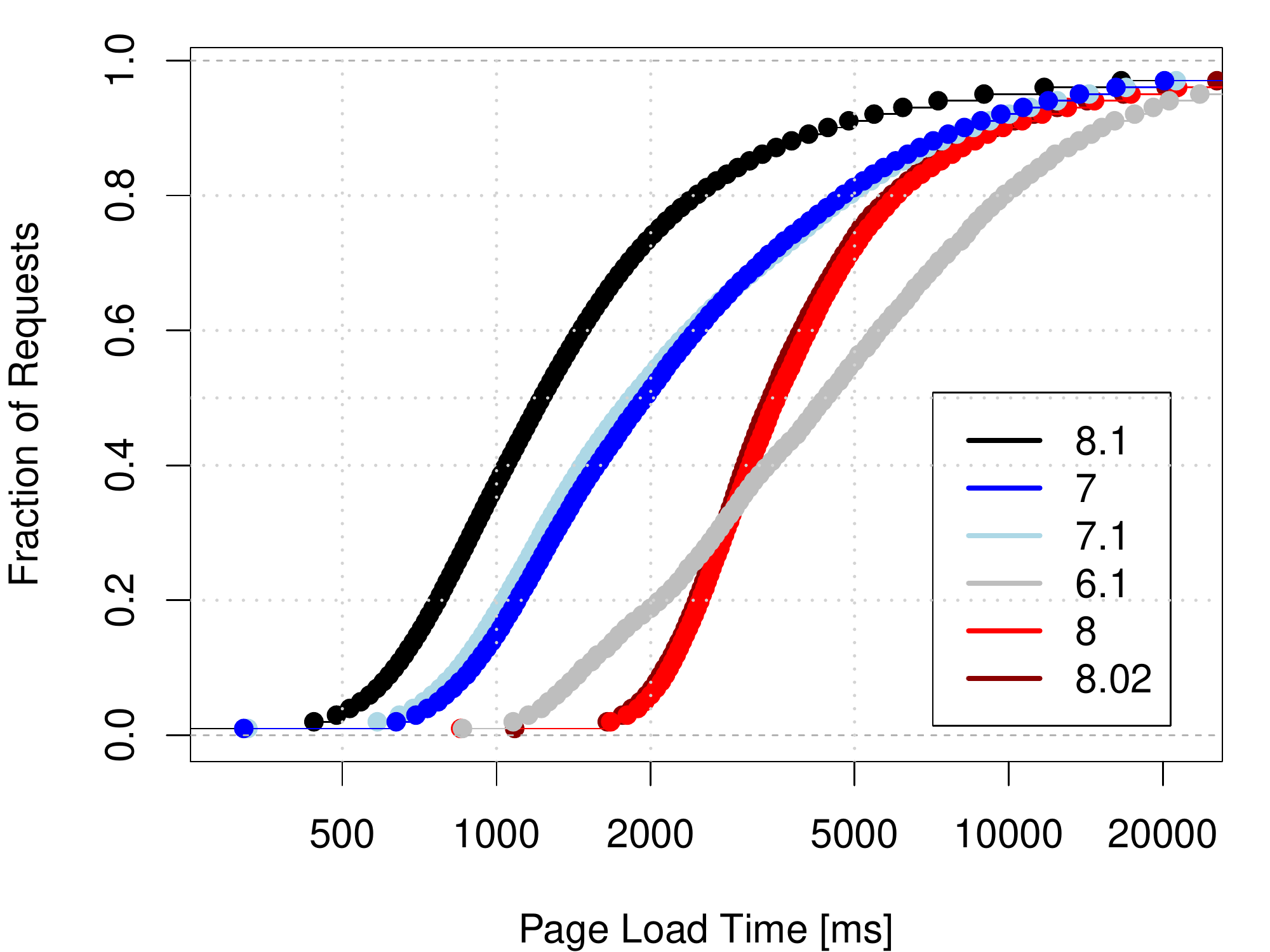}
          \label{fig:iphone}
          }
        \subfigure[Samsung Galaxy]
        {
           \includegraphics[width=\figwidth\textwidth]{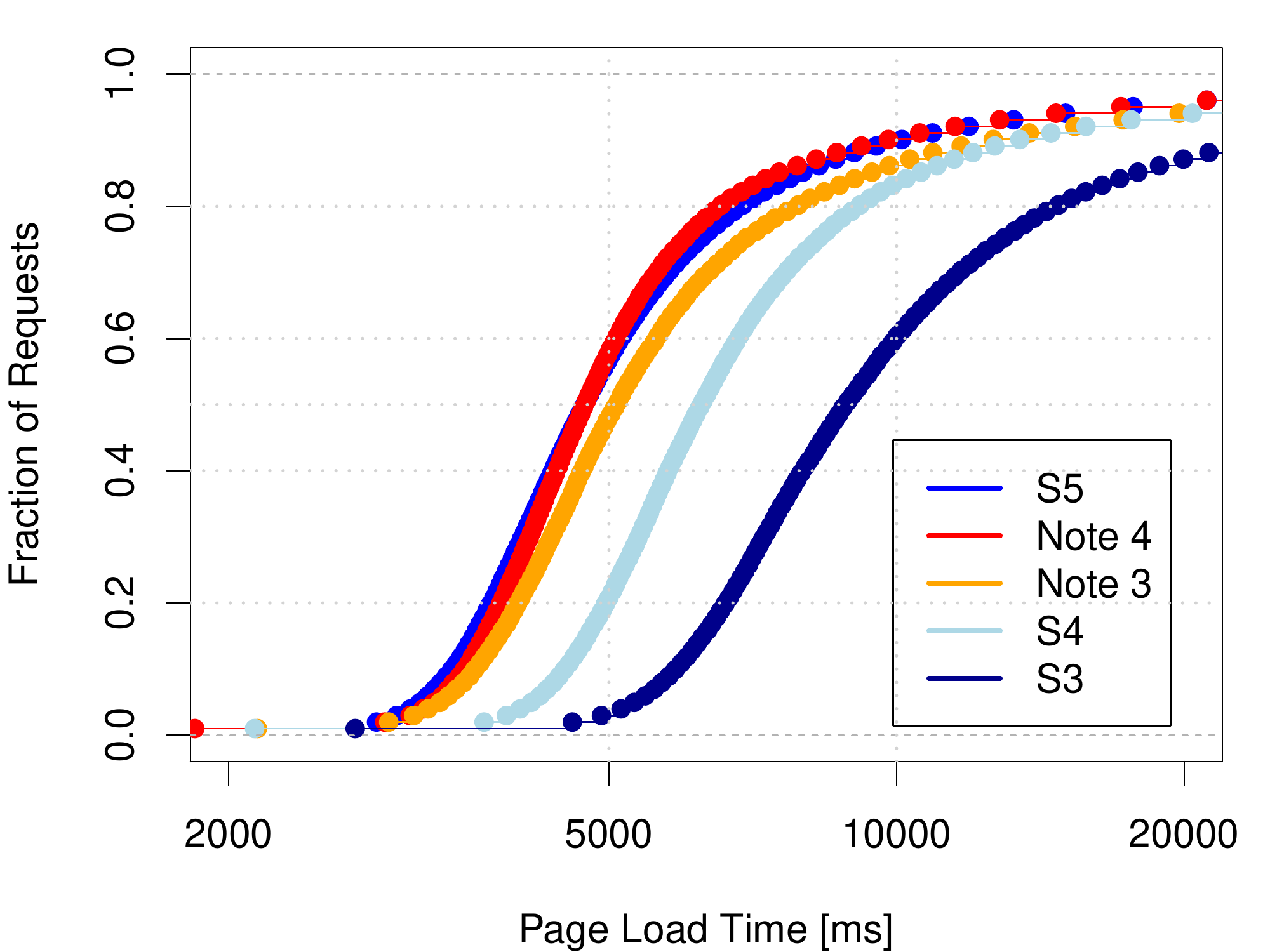}
          \label{fig:samsung}
         }
         \subfigure[HTC]
        {
           \includegraphics[width=\figwidth\textwidth]{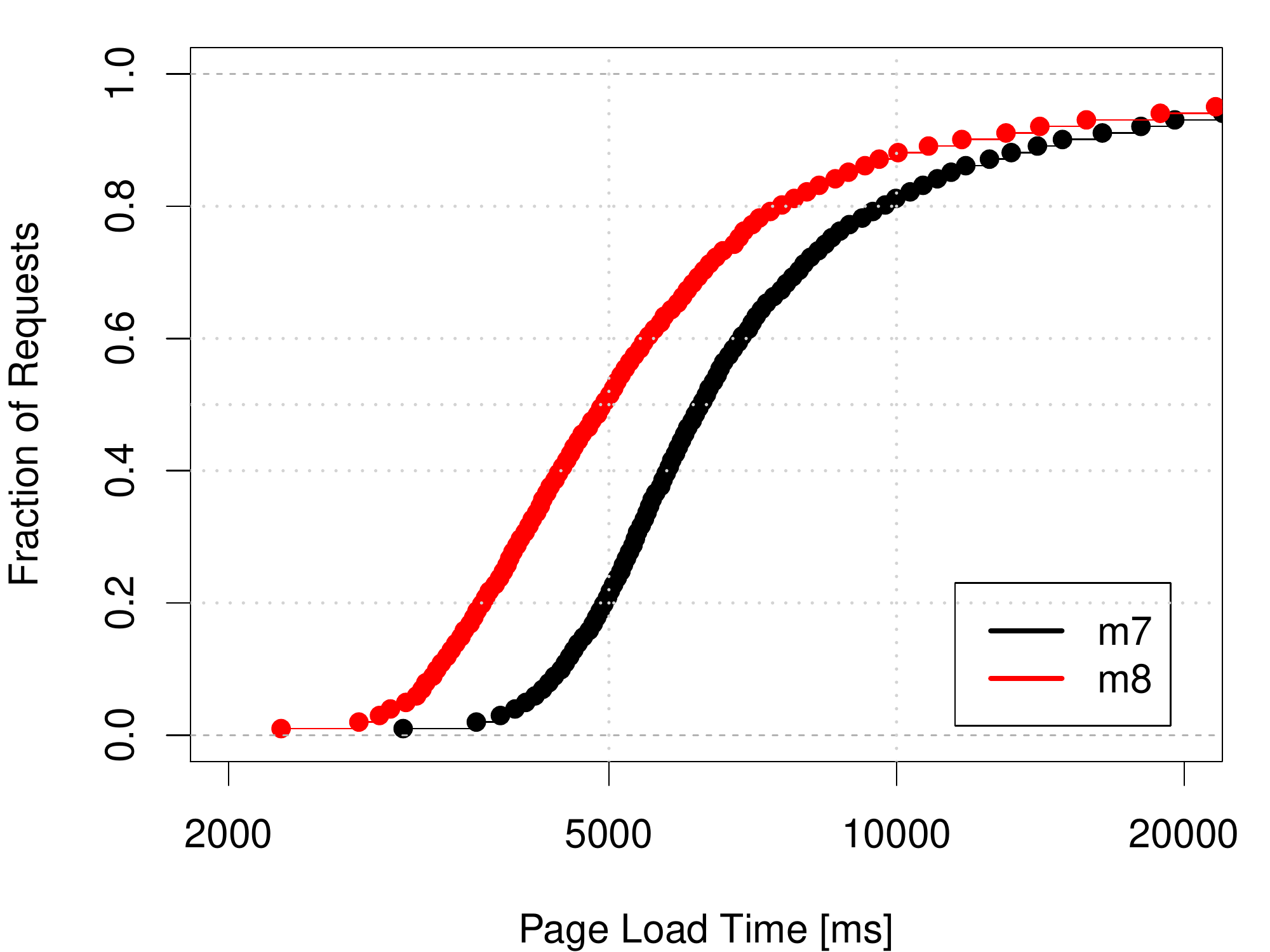}
          \label{fig:htc}
         }
         \subfigure[Nexus]
        {
           \includegraphics[width=\figwidth\textwidth]{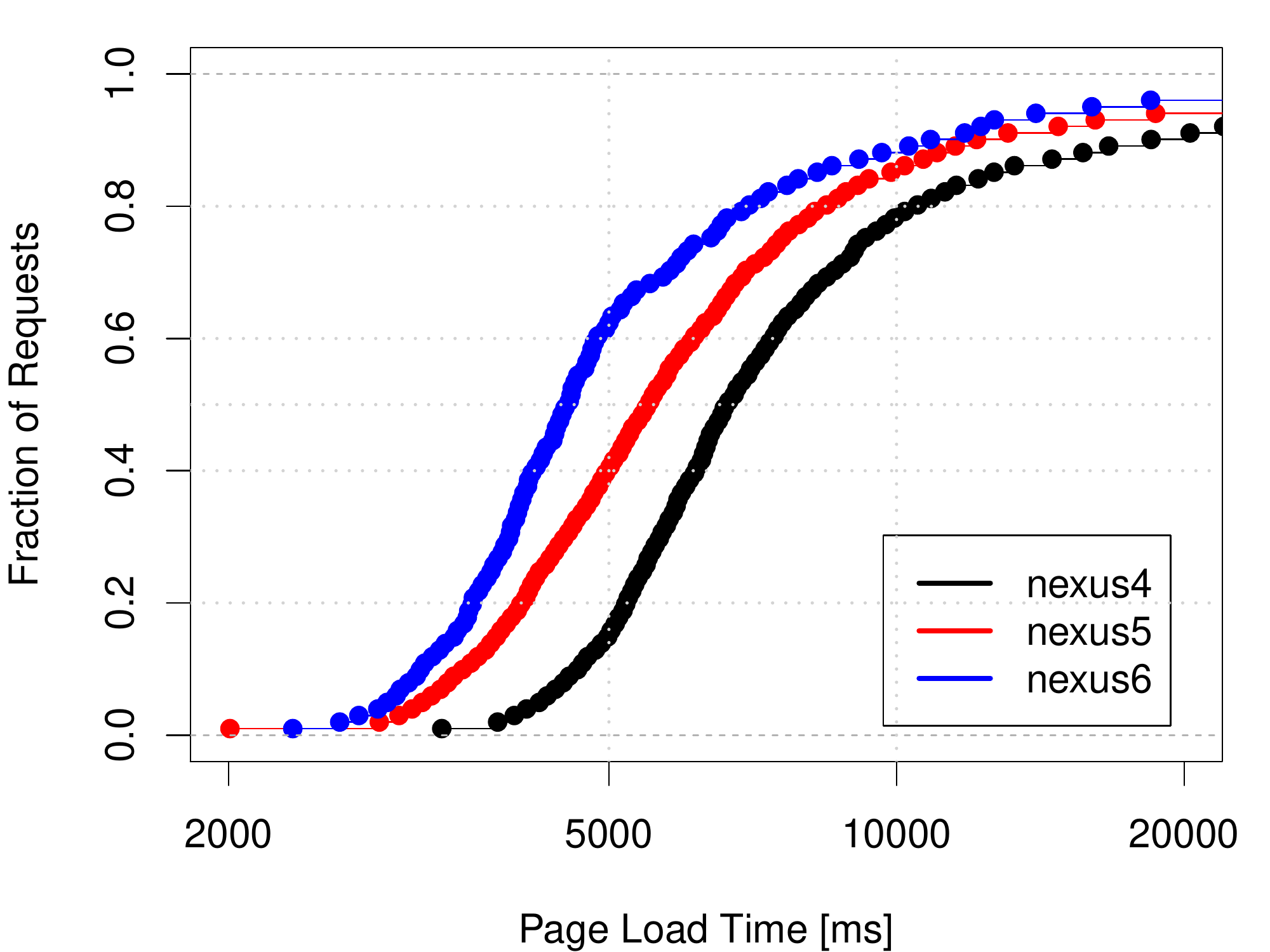}
          \label{fig:nexus}
         }
     \caption{CDFs of the PLTs for different device generations.}
     %\label{fig:mice}
\end{figure*}

The web performance improves greatly with every version of iOS (Figure~\ref{fig:iphone}). The median PLT is reduced by 56\% from version 6.1 to version 7, and by some more 43\% from version 7 to version 8.1. Note that not only the median has been reduced, but that the 90th percentile has been reduced; both are reduced by 71\% from iOS 6.1 to iOS 8.1. 

The versions 8 and 8.02 show a strong regression compared to 7 and 7.1. The upgrade to 8.1 alleviates this issue. The versions 8.2 and 8.3 are very similar to 8.1 in terms of web performance. There are less than 10,000 samples for version 6 and 5.1, this is why those versions are not included in the figure. The median PLT of version 6 is very similar to 6.1; for version 5 we observer a 72\% increase compared to version 6.

To ensure that these findings are not an artifact of the URL chosen, we conducted the same analysis for another URL belonging to a different domain. The data set contains 997,695 samples for this URL from the major cell networks in the US. The median PLT is reduced by 53\% from iOS 6.1 to 7, and by 18\% from 7 to 8.1. The 90th percentile of PLT is reduced by 68\% from iOS 6.1 to 8.1. Those numbers are slightly lower compared to the first URL analyzed, especially the median improvement between version 7 and 8.1 of iOS. 

\subsection{Android devices}

Android devices from various vendors run on Snapdragon chipsets allowing for a straigthforward way to compare the device performance. The core speed typically increases from one device generation to the following one. The data in Table~\ref{table:android}) shows a strong correlation between increasing core speed and a reduction of the PLT. Devices from different vendors running at similar clock speed mostly show very similar web performance.

\begin{table}
\centering
\caption{Web performance of Android devices.}
\label{table:android}
\begin{tabular}{|c|c|r|r|} 
\hline
vendor & model  & frequency (GHz) & median PLT (s) \\
\hline
Samsung & Galaxy S3 & 1.4  & 8.9  \\
Samsung & Galaxy S4 & 1.9 / 1.6 & 6.2  \\
Samsung & Galaxy S5 & 2.5  &  4.7 \\
\hline
Samsung & Note 3 & 2.3 / 1.9 &  5.1 \\
Samsung & Note 4 & 2.7 / 1.9 &  4.7 \\
\hline
HTC & M7 & 1.7 & 6.2  \\
HTC & M8 & 2.3 & 4.9  \\
\hline
Nexus & 4 & 1.5 & 6.7  \\
Nexus & 5 & 2.3 & 5.5  \\
Nexus & 6 & 2.7 & 4.5  \\
\hline
\end{tabular}
\end{table}

\paragraph{Samsung Galaxy}
%Samsung makes it possible to identify which device is used by parsing the user-agent string, e.g. the models \texttt{gt\_i9500} and \texttt{sch\_i545} are both Galaxy S4.
The median PLT has been improved by 30\% from the S3 to the S4~(Figure~\ref{fig:samsung}). The S3 runs a Exynos quad-core processor at 1.4 GHz whereas the S4 exists in multiple versions, running a Snapdragon 600 at 1.9 GHz or a octa-core processor with the faster cores at 1.6 Ghz. The median PLT has further been improved  by 18\% from the S4 to the S5 running a Snapdragon at 2.5 GHz. The Galaxy Note  4 is  on par with the S5, the Note 3 being slightly slower. Both exist in multiple configurations; the faster ones running a Snapdragon 800 at 2.3 Ghz for the Note 3 and a Snapdragon 805 at 2.5 Ghz for the Note 4.

\paragraph{HTC}
The HTC M8 shows a reduction in PLT of 21\% compared to the M7~(Figure~\ref{fig:htc}). A Snapdragon 600  quad-core processor with 1.7 GHz powers the M7, and a Snapdragon 800 (also quad-core) running at 2.3 GHz runs in the M8. 

Compared to the Galaxy series, the M7 is on par with the S4 running at a similar core speed. The M8 is about 5\% slower compared to the S5, both run a Snapdragon 801 chip set but the S5 with 2.5 GHz compared to the 2.3 GHz of the M8.

\paragraph{Nexus}
The Nexus 4 and 5 are produced by LG, and the Nexus 6 is produced by Motorola. The web performance has been greatly improved from the version 4 to version 5; the median PLT is reduced by 18\% (Figure~\ref{fig:nexus}). The processor has been updated from a Snapdragon S4 1.5 GHz pro to a Snapdragon 800 running at 2.3 GHz (both quad-core). Another 18\% reduction is observed switching from the Nexus 5 to the Nexus 6 which runs a Snapdragon 805 at 2.7 GHz.

The HTC M7 and the Galaxy S4 are 6\% faster than the Nexus 4, both of them running cores with a slightly higer clock speed than the S4. The Nexus 5 looses some ground, being 10\% slower than the M8 (same clock speed) and 14\% slower than the S5 with a 200 MHz faster processor. The Nexus 6 is the fastest phone in our analysis with a browser implementing Navigation Timing being the only phone to run at 2.7 GHz.

\section{Impact of the network performance on the Page Load Time}
\label{sec:network}

When comparing the impact of the networks used on the PLT we need to make sure that no bias is introduced. One network operator might promote high-end device while another operator is focusing on the market for more affordable smartphones. Different demographics might also use different networks and browse different content. To exclude possible biases, we limit the data set to the same single URL used in the device section~(Sec.~\ref{sec:device}) and also consider only one single device.

\begin{figure*}[ht]
     \centering
       \subfigure[Four major cell networks in the US and one fiber to the home network with wifi (in black).]
        {
           \includegraphics[width=\figwidthtriple\textwidth]{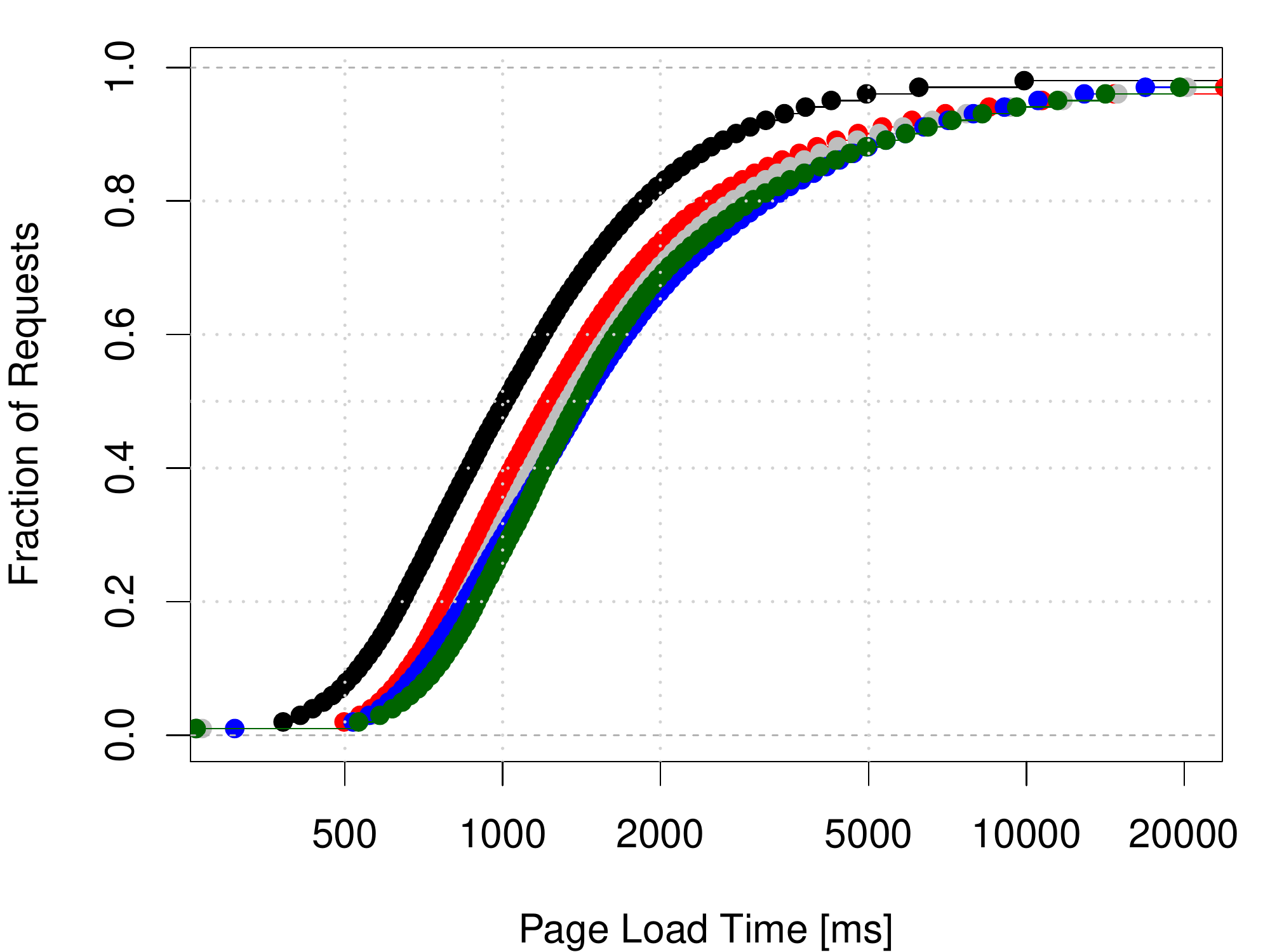}
          \label{fig:network-us}
          }
      \subfigure[Four major cell networks in France and one DSL network with wifi (in black).]
        {
           \includegraphics[width=\figwidthtriple\textwidth]{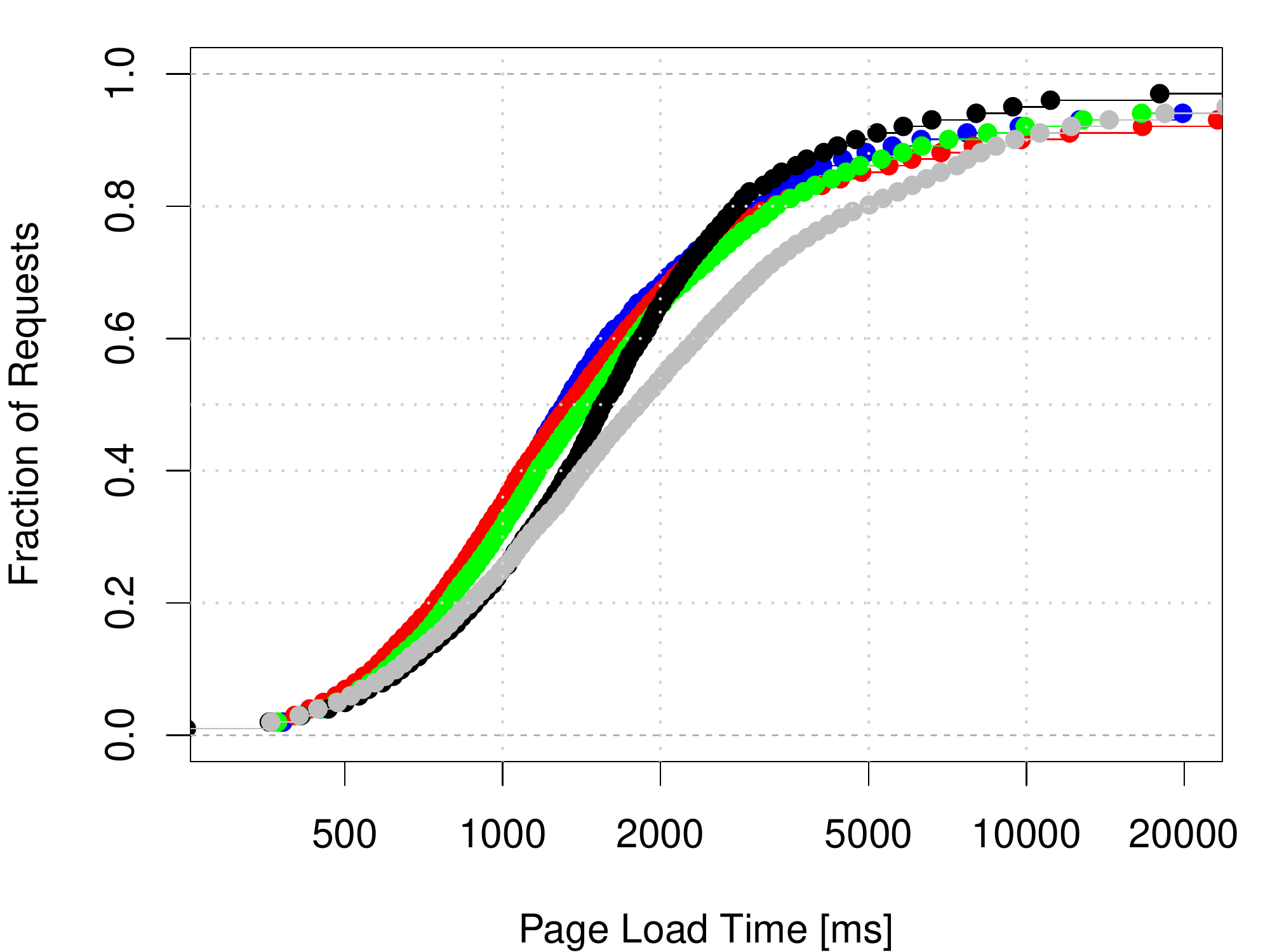}
          \label{fig:network-fr}
         }
      \subfigure[Cellular networks in selected countries.]
        {
           \includegraphics[width=\figwidthtriple\textwidth]{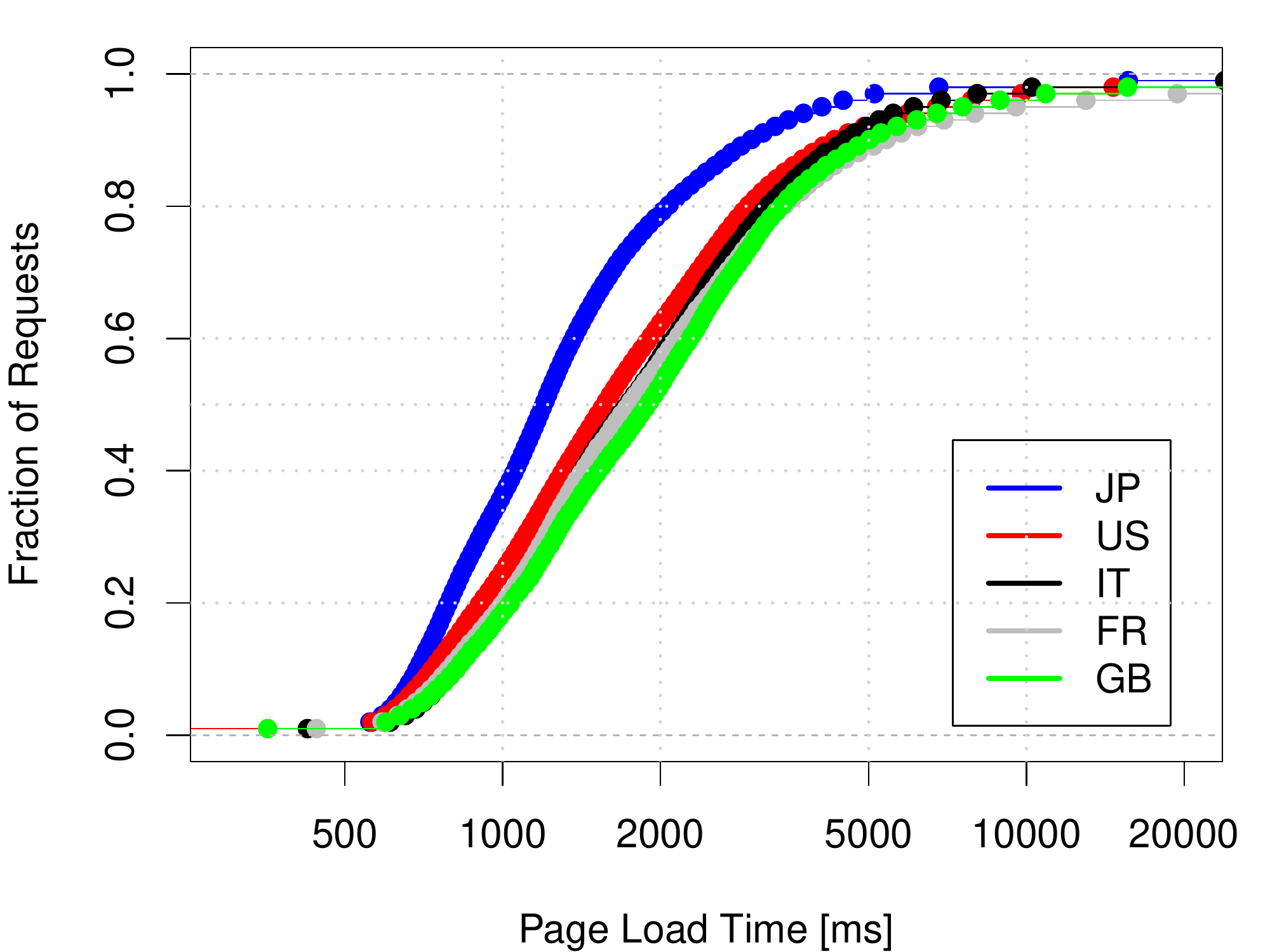}
          \label{fig:countries}
         }
     \caption{CDF of the PLTs experienced by mobile devices on various cellular networks.}
     %\label{fig:mice}
\end{figure*}

Four major cellular networks in the US are compared against each other and against a FTTH network using wifi for the last meters~(Figure~\ref{fig:network-us}). The shapes of the distributions are very similar for all networks, including the FTTH network. Samples from only one URL and only one device (iOS 8.1) are used for this analysis; 500,135 samples for all networks combined. The fastest cellular network shows a median PLT 14\% faster than the slowest cellular network. Users connected on wifi backed by a FTTH provider experience median PLTs of 18\% faster than the fastest cellular network and even 28\% faster than the slowest cellular network in this set.

Repeating the same exercise for another country (France) shows a similar pattern~(Figure~\ref{fig:network-fr}). The major cellular networks have comparable PLTs within 9\% percent of one another, with an exception for one network that has a median PLT 38\% slower than the fastest network. The shape of the distribution is similar for all four networks. The figure also shows the distribution for one home network with DSL access technology plus wifi for the last few meters (black in the figure). The median performance is slightly worse than the three well-performing cell networks (19\% slower than the fastest cell network), and slightly better than the slower network. The shape of the distribution is different than that of one of the cell networks, the fastest 25\% of the samples being as slow as the slowest cell network, whereas the slowest 25\% are faster than the fastest cell network. Samples from only one URL and only one device (iOS 8.1) are used for this analysis; 30,287 samples in total.

These numbers show that cellular networks have come a long way. Only a few years ago mobile browsing required a lot of patience, today the mobile performance is comparable to what people are experiencing at home. The user experience provided by a cellular network and wired home  network (plus wifi) are similar.

In order to compare the web PLT on cellular networks in different countries we had to find a single URL that has a significant number of hits in those countries. This turned out to be harder than imagined; most sites show a highly regionalized access pattern. Figure~\ref{fig:countries} plots the CDF of PLTs for devices with iOS 8.1 and for one URL with at least several tens of thousands of samples for each country (297,039 samples in total). In the fastest country, Japan, the median time is 37\% smaller compared to the slowest country, Great Britain. The remaining countries, France, Italy, and the US, show a very similar performance with a slight advantage for the US.

\section{Discussion}
\label{sec:discussion}
Navigation Timing helps to obtain a picture of web performance in general. Issues with specific devices, networks, countries, or sites can be isolated. Analyzing the different stages, such as DNS look up time, connect time, time to first byte, Document Object Model (DOM) complete, and finally the load event being fired, it is possible to point to specific problems. It might take a long time to connect to the server; or once the connection is established it takes a long time for the first bytes to reach the client; or the period after the DOM has been built and before the load event fires is longer than expected. Best practices have been established to address well-known issues with web performance~\cite{grigorik}.

Navigation Timing makes it simpler to collect timing metrics for web pages, however it does not reveal information about individual resources on those pages. The recent Resource Timing interface provides similar information to Navigation Timing at a resource level~\cite{restiming}. Waterfall charts (also present in various web browser developer tools) providing a clear visualization of when each resource on a page is loaded can be constructed from the obtained data. They show in detail how resources interact and which resources have the most performance impact on PLTs.

\section{Related Work}
\label{sec:relwork}
The advanatages of collecting performance data from real users with the Navigation Timing API as opposed to perform very limited tests from only a few vantage points are outlined in~\cite{meenan2013fast}. The author discusses the various metrics provided by the API and shows an example on how to gain insights into the performance experienced by group of users by analyzing the collected data.
Wprof~\cite{wang2013wprof} is a tool that analyzes the dependencies of objects on a website and computes the critical path of the page load time. The authors ran tests for popular websites from their campus network using one machine (3 Ghz clock speed) and find that for their measurement setup computing time accounts for 35\% of the median page load time and network related time for the remainder. Additional to main results the experirement was repeated using two different networks and with two other machines with lower computing power. The networks did not have a significant effect on the fraction of time spent for computation and network transfer. The experiment with a 2 Ghz machine showed that the time spent with computation increased to 40\%. These results are very relevant to our study since they already show a correlation between computing speed and web performance.
A very recent study~\cite{nikravesh2015mobilyzer} (published after the data collection for this paper had been completed) presents a platform to run mobile network measurement experiments. It presents crowd-sourced data collected via the Navigation Timing API by 80 users and conclude that web performance can significantly benefit from faster processors. Unlike our study the effect of the network performance is not considered, in fact even wifi and cellular networks are mixed together in the results.

%http://queue.acm.org/detail.cfm?id=2446236

% \url{http://www.webpagetest.org/}

% http://www.stevesouders.com/blog/2010/05/07/wpo-web-performance-optimization/

% http://analytics.blogspot.com/2012/03/measure-your-websites-performance-with.html

%Mobilyzer:  \cite{nikravesh2015mobilyzer}

\section{Conclusion}
\label{sec:conclusion}
In this paper we have analyzed the impact of device performance and network performance on the PLT. The methodology used consists of inserting javascript into a small sample of websites, leveraging the Navigation Timing interface (when available) to collect navigation timing. In a subsequent step this data is collected in a central data store and enriched with device and network specific information. By querying the data store PLTs of devices, we can compare network, or sites, when the two other factors are isolated. 

Thanks to a unique data set provided by a large CDN this study shows that upgrading the device from one generation to the next has a significant impact on the PLT distribution. The performance differences between well-deployed LTE networks, on the other hand, are not as significant in this study, compared to the difference between upgrading the device. The performance gap between cellular networks to a FTTH network in the US is less pronounced than the gap between device generations. In France, home networks using DSL are even slower at loading pages than most cellular networks.

Given the performance gap between smartphone generations, upgrading regularly may be advisable for the performance-minded consumer.

Opinions or conclusions expressed this paper, if any, are not necessarily those of Akamai; Akamai assumes no liability for use of, or reliance on, any information provided.


%\small
\bibliographystyle{abbrv}
\bibliography{rumpaper}  % sigproc.bib is the name of the Bibliography in this case

\begin{thebibliography}{1}

\bibitem{navtiming}
Navigation timing specification.
\newblock
  \url{https://dvcs.w3.org/hg/webperf/raw-file/tip/specs/NavigationTiming/Overview.html}.

\bibitem{restiming}
Resource timing working draft.
\newblock \url{http://www.w3.org/TR/resource-timing/}.

\bibitem{userstudy}
Small screens mighty sales.
\newblock
  \url{https://www.internetretailer.com/2015/09/01/small-screens-mighty-sales}.

\bibitem{grigorik}
I.~Grigorik.
\newblock {\em High Performance Browser Networking: What every web developer
  should know about networking and web performance}.
\newblock O'Reilly Media, 2013.

\bibitem{meenan2013fast}
P.~Meenan.
\newblock How fast is your website?
\newblock {\em Communications of the ACM}, 56(4):49--55, 2013.

\bibitem{nikravesh2014mobile}
A.~Nikravesh, D.~R. Choffnes, E.~Katz-Bassett, Z.~M. Mao, and M.~Welsh.
\newblock Mobile network performance from user devices: A longitudinal,
  multidimensional analysis.
\newblock In {\em Proceedings of PAM}, pages 12--22, 2014.

\bibitem{nikravesh2015mobilyzer}
A.~Nikravesh, H.~Yao, S.~Xu, D.~Choffnes, and Z.~M. Mao.
\newblock Mobilyzer: An open platform for controllable mobile network
  measurements.
\newblock In {\em Proceedings of MobiSys}, pages 389--404, 2015.

\bibitem{wang2013wprof}
X.~S. Wang, A.~Balasubramanian, A.~Krishnamurthy, and D.~Wetherall.
\newblock Demystifying page load performance with wprof.
\newblock In {\em Proceedings of NSDI}, pages 473--485, 2013.

\bibitem{wang2011web}
Z.~Wang, F.~X. Lin, L.~Zhong, and M.~Chishtie.
\newblock Why are web browsers slow on smartphones?
\newblock In {\em Proceedings of HotMobile}, pages 91--96, 2011.

\end{thebibliography}
% ACM needs 'a single self-contained file'!
%

% That's all folks!
\end{document}